\documentclass[aps,twocolumn,prl]{revtex4}
\usepackage{newlfont}
\usepackage{amssymb}
\usepackage{amsfonts}
\usepackage{amsmath}
\usepackage{wasysym}
\usepackage{graphicx}
\usepackage{bm}

\usepackage{epsfig}

\usepackage{amsthm}

\begin{document}

\title{Conditions for Monogamy of Quantum Correlation: \\Greenberger-Horne-Zeilinger versus W states}
%\title{Conditions for Monogamy of Quantum Correlation:\\ Monogamous Greenberger-Horne-Zeilinger versus Polygamous W states}
%\title{Conditions for Monogamy of Quantum Discord:\\ Monogamous Greenberger-Horne-Zeilinger versus Polygamous W states}
%\title{Conditions for Monogamy of Quantum Correlations:\\ Monogamous Greenberger-Horne-Zeilinger versus Polygamous W states}
%\title{Conditions for Monogamy of Quantum Correlations:\\ Polygamous W vs. Monogamous GHZ}
%\title{Conditions of Monogamy of Quantum Correlations: Polygamous W vs. Monogamous GHZ states}
%\title{Monogamy of Quantum Correlations as Detector of Genuine Tripartite Entangled States}

\author{R. Prabhu, Arun Kumar Pati, Aditi Sen(De), and Ujjwal Sen}

\affiliation{Harish-Chandra Research Institute, Chhatnag Road, Jhunsi, Allahabad 211 019, India}

\begin{abstract}
%In the quantum world, q
Quantum correlations  are expected to respect all the conditions required 
for them to be good measures of quantumness in the bipartite scenario.
%follow all the necessary conditions required for them 
%to be ``good'' quantum correlation measures. 
In a multipartite setting, sharing entanglement between several parties is restricted by the 
monogamy of entanglement. 
%It tells that if two parties are highly entangled, then they can not be highly entangled with any other third party simultaneously. 
We take over the concept of monogamy to an information-theoretic quantum correlation measure, and find that it violates monogamy in general. 
Using the notion of interaction information, 
we identify necessary and sufficient conditions for the measure to obey monogamy, for arbitrary pure and mixed quantum states. 
%As an application, w
We show that while three-qubit generalized 
Greenberger-Horne-Zeilinger states follow monogamy, generalized W states do not. 
%The results have further potential applications in quantum communication devices. 
%The monogamy of discord will give another tool to detect two
% inequivalent class of states.  

\end{abstract}

\maketitle

\emph{Introduction and main results.--} Quantifying quantum correlations plays an important role 
in the development of quantum information science.
%  
%for explaining several quantum mechanical phenomena \cite{ekhane-NC}. 
In particular, entanglement 
%as a measure of quantum correlations 
%is a physical property that 
has been successfully employed to interpret several phenomena which cannot be understood within 
%clarified 
 classical physics \cite{HHHH-RMP}. It has also been identified as the basic ingredient for different quantum communication protocols \cite{comm-review}
% like 
%super-dense coding \cite{dc}, teleportation \cite{tele}, quantum cryptography \cite{Ekert91}, remote-state preparation \cite{remote}
and quantum computational tasks 
%such as the one-way quantum computer 
\cite{Briegel}.
Moreover, bipartite as well as multipartite entangled states are prepared and are being successfully used to 
implement different quantum information protocols.
% in several laboratory around the globe 
%\cite{experiments}. 

However, in the past, several quantum phenomena have been discovered in which  entanglement is absent.
 These include ``quantum nonlocality without entanglement''  -- locally indistinguishable orthogonal product states
\cite{nlwe}, and the model of deterministic quantum computation with one quantum bit 
%\textbf{a model with a sufficently
% low entanglement which can lead to exponential speed-up over its classical counterpart} 
\cite{KnillLaflamme,    Animesh, others}.
%, separable states
% with nonclassical features, performing quantum information processing tasks better than any classical protocols \cite{others}.
%
%
In view of this, it is important and interesting 
%There are several attempts 
to quantify quantum correlations in a multiparticle quantum state, independent of the entanglement-separability paradigm.
% from which are different from entanglement. 
An important 
attempt in this direction is the introduction of 
%
%example of such an attempt is the 
quantum discord \cite{discord1, discord2}  for bipartite systems. Discord is an information-theoretic physical quantity 
that characterizes quantum correlations. This arises because 
% by ``discord'' in the quantum level 
 two classically equivalent definitions of mutual information do not match in the quantum level. Recently, the concept of quantum discord has been 
generalized to the multipartite scenario using e.g. the notions of dissonance \cite{Modi} and dissension \cite{Arun-disso} 
(see also \cite{discord-multi,workdeficit, others, bakisob}). 
%Other works in this direction include \cite{workdeficit, others, bakisob}.
%Among them, the information theoretic measure of 
%quantum correlations -- quantum discord \cite{discord1, discord2} (cf. \cite{workdeficit}) was used to characterize nonclassicality in the above
% scenarios \cite{others, bakisob}. 

%All the quantum correlation measures should satisfy few reasonable properties required to be a ``good measure''  -- (i) they 
%require to be vanishing for  product states, (ii) nonincreasing under local operations and classical communication (LOCC). All the entanglement 
%measures \cite{HHHH-RMP} and discord for bipartite systems satisfy these criterion. On top of that, in a multipartite scanario, entanglement is monogamous -- if 
%two parties are highly entangled, these two parties cannot share more entanglement with other parties \cite{Wootters, monogamyN, monogamy}. 

An important property that is satisfied by entanglement in multiparty quantum states is that it is monogamous -- if 
two subsystems are highly entangled, then they cannot share a substantial amount of entanglement with other subsystems \cite{Wootters, monogamyN}. 
In this paper, we ask whether such monogamous character is an intrinsic property of other quantum correlation measures, in particular, 
of quantum discord. 
We carry over the monogamy concept beyond the paradigm of entanglement, and 
ask:
% the following question: 
\emph{Does the sharing of general quantum correlations follow the same broad guidelines that are followed by entanglement?
In particular, does quantum discord satisfy the monogamy relation?}
We find that the answers in general are negative. 
For another instance
where monogamy considerations have been taken to an
information-theoretic quantum multiparticle measure, see Ref. \cite{natun}.
%Quantum discord can be polygamous. 
Using the concept of interaction information, we identify necessary as well as sufficient 
%simplified 
criteria 
for quantum discord to be monogamous. 
The concept of interaction information in the context of classical information theory is an important one and it has 
many applications in various disciplines, including biophysics, medicine, data analysis, classical control theory, and Bell inequalities \cite{int}. 
%In information theory, it is usually regarded as unwelcome, as it can be both positive and negative. 
Here, we find that it naturally fits  in the context of monogamy of quantum correlation, and 
it explains as to when it can be satisfied and violated.  
Interestingly, we find that the monogamy condition can be applied to distinguish two inequivalent classes
of quantum states in a three-party situation. Specifically,  we show that the generalized 
Greenberger-Horne-Zeilinger (GHZ)  states \cite{GHZ} always satisfy the monogamy 
of quantum discord, while the  
generalized  W states \cite{Wstate, dur-vidal-cirac}  always violate the same. This therefore gives us a 
physically interesting method to detect the generalized GHZ against generalized W states.
%: if a supplier selling quantum states promises to provide either generalized GHZ or generalized W, we can use the monogamy condition to find out what the state is.
%The space of all t
Three-qubit pure states can be divided into the GHZ and  W classes, of which the generalized GHZ and W are 
respective proper subsets. Members of the two classes 
%Any member state of any of these two classes 
cannot be transformed into each other by local quantum operations and classical communication (LOCC),
%into any member state of the other class, 
with a nonzero probability of success \cite{dur-vidal-cirac}. This has subsequently been applied in a wide variety of areas ranging from quantum information to black holes  \cite{ilahabad-station-er-baire}. 
In course of our 
investigation, numerical searches in the GHZ and W classes have revealed 
that members of the GHZ class may or may not violate the monogamy relation, while those in the W class are always violating. 

%It is plausible that the states of the W class would always violate monogamy, whereby 
%we would have a witness for GHZ class states by using the monogamy condition. 
%
%is satisfied the state is sure to be a generalized 
%GHZ, provided only generalized GHZ and generalized W were the options. 
%satisfy monogamy of discord
% while . 

%The paper is organized as follows: In Sec. II, we will give a short introduction to quantum discord.
%A discussion on monogamy will follow in Sec. III. 
%Necessary and sufficient conditions for a state to satisfy monogamy of quantum discord will be presented in Sec. IV. 
%Sec V will be devoted to a demonstration to indicate that the  two inequivalent 
%classes of three-qubit pure states behave differently with respect to the monogamy relation for discord. 
%We summarize our results in Sec. VI. 

\emph{Measure of quantum correlation.--}
%: quantum discord.}
 Quantum discord \cite{discord1, discord2} is defined as the difference between two classically equivalent expressions for the mutual information, when 
extended to the quantum regime:
\(D(\rho_{AB})= \tilde{I}(\rho_{AB}) - I(\rho_{AB})
%\nonumber \\
            = S(\rho_{A|B}) - \tilde{S}(\rho_{A|B})\).
(In \cite{discord1}, \(\tilde{I}\) and \(I\) are respectively denoted as \(I\) and \(J\).)
Here \(\tilde{I}(\rho_{AB})= 
%S(\rho_A)+ S(\rho_B)- S(\rho_{AB})
%\nonumber\\
 %           \equiv 
 S(\rho_A) - \tilde{S}(\rho_{A|B})\),
%\end{eqnarray}
is the quantum mutual information, and can be interpreted as the total correlations in \(\rho_{AB}\), 
where \(\rho_A\) and \(\rho_B\) are the local density matrices of \(\rho_{AB}\). 
\(S(\sigma) = -\mbox{tr}(\sigma \log_2 \sigma)\) is the von Neumann entropy of a density matrix \(\sigma\). 
 \(\tilde{S}(\rho_{A|B}) = S(\rho_{AB}) - S(\rho_B)\) is the ``unmeasured'' quantum conditional entropy \cite{qmi} (see also \cite{Cerf, GROIS})). 
On the other hand,  \(I(\rho_{AB}) = S(\rho_A) - S(\rho_{A|B})\) can be interpreted as the classical correlations in \(\rho_{AB}\),
%\end{equation}
where the quantum conditional entropy is defined as  
%\begin{equation}
%\label{borphi}
\(S(\rho_{A|B}) = \min_{\{\Pi_i^B\}} \sum_i p_i S(\rho_{A|i})\),
 %\end{equation}
with the minimization being over all 
%Here we assume that the 
projection-valued measurements, \(\{\Pi^B_i\}\),  performed on subsystem \(B\).
% using the set of projectors , 
Here \(p_i = \mbox{tr}_{AB}(\mathbb{I}_A \otimes \Pi^B_i \rho_{AB} \mathbb{I}_A \otimes \Pi^B_i)\) is the probability of obtaining the outcome \(i\), and 
the corresponding post-measurement state  
%and the ensemble produced 
for the subsystem \(A\) is \(\rho_{A|i} = \frac{1}{p_i} \mbox{tr}_B(\mathbb{I}_A \otimes \Pi^B_i \rho \mathbb{I}_A \otimes \Pi^B_i)\), 
where \(\mathbb{I}_A\) is the identity operator on the Hilbert space of the quantum system that is in possession of \(A\).
A nonzero quantum discord  signifies the presence of quantum correlations in a bipartite quantum state.

\emph{Monogamy.--}
Monogamy of quantum correlations is a property satisfied by certain entanglement measures in a 
%quantum a concept in a 
multipartite scenario. Given a multipartite state \(\rho_{A_1 A_2\ldots A_N}\) shared between \(N\) parties, 
 the monogamy condition for a bipartite quantum correlation measure \({\cal Q}\) 
assures that the bipartite quantum correlations in the multiparty state are distributed in such a way that the following 
relation is satisfied: 
%the distribution of bipartite quantum correlation amongst different parties which is
%\begin{equation}
\({\cal Q}(\rho_{A_1 A_2}) + {\cal Q}(\rho_{A_1 A_3}) + \cdots + {\cal Q}(\rho_{A_1 A_N}) \leq {\cal Q}(\rho_{A_1: A_2 A_3 \ldots A_N})\),
%\end{equation}
where \(\rho_{A_1A_2} = \mbox{tr}_{A_3 A_4 \ldots A_N}\rho_{A_1A_2 \ldots A_N}\), etc. 
%\({\cal Q}\) is any measure of quantum correlations, which is considered among all bipartite reduced density matrices of the \(N\)-party state
%\(\rho_{A_1 A_2\cdots A_N}\) and 
On the RHS, the quantity  \({\cal Q}(\rho_{A_1: A_2 A_3 \ldots A_N})\) is the same quantum correlation measure obtained in the bipartition \(A_1: A_2 A_3 \ldots A_N\)
of the \(N\) parties.
It was shown that certain entanglement measures 
%called tangle 
satisfy the above monogamy inequality \cite{Wootters, monogamyN}. In this paper, we ask whether quantum discord 
satisfy the  same relation. 
%carry over this concept beyond the paradigm of entanglement, and 
%ask:\\
%% the following question: 
%\emph{Does quantum discord satisfy this monogamy relation? Does the sharing of quantum discord follow the same broad guidelines that are followed by entanglement?}\\
In  particular, 
%we will answer this question for the tripartite scenario, i.e., 
we will consider whether quantum discord of a tripartite state \(\rho_{ABC}\) 
satisfy the following inequality:
\begin{equation}
\label{monogamydiscord}
% \label{aaj-1stquicdinner}
 D(\rho_{AB}) + D(\rho_{AC}) \leq D(\rho_{A: BC}). 
\end{equation}
Violation of the above inequality will imply that discord is polygamous for the corresponding state.

\emph{Necessary and sufficient criteria for quantum discord to be monogamous.--}
%In this section, we 
We will now present  the conditions that signal whether a multi-site quantum state is monogamous in nature with respect to quantum discord. 
We begin with Theorem I, which will tell us as to when the quantum correlation will
respect monogamy.
%, while Theorem II will tell us when it will be polygamous. 
%Then we will show that refuting Theorem I can lead to polygamous behavior.
Before proving the results, we need to present a few definitions about mutual information, 
conditional mutual information, and interaction information.

\noindent {\bf Definition I:} 
%The mutual information for a bipartite state
%$\rho_{AB} $ is $I(A:B) = S(A) + S(B) - S(A,B)= S(A) - S(A|B)$ 
%and 
The mutual information and conditional entropy for the two-party cases have already been defined. 
For a tripartite state \(\rho_{ABC}\),  the unmeasured conditional mutual 
information \(\tilde{I}(\rho_{A:B|C}) = \tilde{S}(\rho_{A|C}) - \tilde{S}(\rho_{A|BC})\), and the interrogated conditional mutual information 
\(I(\rho_{A:B|C}) = S(\rho_{A|C}) - S(\rho_{A|BC})\).
%
%, for a tripartite state \(\rho_{ABC}\) is \(I(\rho_{A:B|C}) = {\cal S}(\rho_{A|C}) - {\cal S}(\rho_{A|BC})\),
%where \({\cal S}(\rho_{A|C})\) can be 
%\begin{itemize}
%\item 
%\(\tilde{S}(\rho_{A|C})\),
%, as defined in (\ref{qmi1}), 
%or 
%\item 
%\(S(\rho_{A|C})\), 
%as defined in (\ref{borphi}), 
%or
%\item 
%\(S(\rho_{A|C})_{\{\Pi_i^C\}}\), which is the quantum conditional entropy where the measurement is performed by using the 
%\emph{specific} 
%complete measurement \(\{\Pi_i^C\}\) on the subsystem \(C\). 
%\end{itemize}
% (as defined in (\ref{})). 
%where $S(A|B), S(A|C)$ and $S(A|B, C)$ are the conditional entropies.
Here, 
%\(I(\rho_{AB})\), 
\(\tilde{I}(\rho_{A:B|C})\) and  \(I(\rho_{A:B|C})\) are nonnegative, which  follow from the 
fact that 
\(S(\rho_{A|C}) \ge S(\rho_{A|B C})\) and \(\tilde{S}(\rho_{A|C}) \ge \tilde{S}(\rho_{A|B C})\), which say that conditional entropy is nonincreasing 
when conditioned on more parties.

%\noindent {\bf Definition II:} For a tripartite system, the interaction information \cite{eitaCoverThomas}
%is defined as \(I(\rho_{ABC}) = I(\rho_{A:B|C}) - I(\rho_{AB})\). 

%Even though the conditional mutual information and the mutual information 
%are both positive, the interaction information can be either positive 
%or negative.

\noindent {\bf Definition II:} We carry over the concept of interaction information from classical information 
theory \cite{eitaCoverThomas} to quantum mechanics and define the (unmeasured) interaction information, $\tilde{I}(\rho_{ABC})$, as  
%\[
%\begin{eqnarray}
\(\tilde{I}(\rho_{ABC}) = \tilde{I}(\rho_{A:B|C}) - \tilde{I}(\rho_{AB}) 
%\nonumber \\ 
%\nonumber\\
%
%as the one when 
%we do not perform any measurement on subsystems (i.e., conditional entropies do not
%involve measured density operators). We call this the unmeasured interaction 
%information of the tripartite quantum state, and is given by 
%%\begin{eqnarray}
%\(I(\rho_{ABC}) = I(\rho_{A:B|C}) - I(\rho_{AB})
%\nonumber \\
%&=& \left[\tilde{S}(\rho_{A|C}) - \tilde{S}(\rho_{A|BC})\right] - \tilde{I}(\rho_{AB})  \nonumber\\
%\end{eqnarray}
%\begin{eqnarray}
%\label{eikhane-ek}\\
%&=& 
= S(\rho_{AB}) + S(\rho_{BC}) + S(\rho_{AC}) 
% \nonumber\\
-\left[S(\rho_{A}) + S(\rho_{B}) + S(\rho_{C})\right] - S(\rho_{ABC})\).
%\,
%\end{eqnarray}
%\].
%\,\, \,\,\label{eikhane-dui}
%\end{eqnarray}
Note here that since this interaction information is without ``interrogation'' (measurement), the conditional entropies 
%in Eq. (\ref{eikhane-ek}) 
are defined here as 
\(\tilde{S}(\rho_{A|C}) = S(\rho_{AC}) - S(\rho_C)\) and 
\(\tilde{S}(\rho_{A|BC}) = S(\rho_{ABC}) - S(\rho_{BC})\), so that no measurement is required for their definitions.
%This is obtained from the definition of interaction information, in which for the definitions of conditional information, we use 
%\(S()\) 
%
%
Interrogated interaction information is defined by using  
conditional entropies that involve density operators of one subsystem given that a  
complete measurement has been performed  on another subsystem. For the state \(\rho_{ABC}\), and a given set of measurements, an interrogated interaction information is given by
%\begin{equation}
%\label{raat2-to-chhoi}
\(I(\rho_{ABC})_{\{\Pi_k^B, \Pi_i^C,\Pi_j^{BC}\}}  = I(\rho_{A:B|C})_{\{\Pi_i^C,\Pi_j^{BC}\}}  - I(\rho_{AB})_{\{\Pi_k^B\}}\),
%\end{equation}
where the suffix on \(I(\rho_{ABC})_{\{\Pi_k^B, \Pi_i^C,\Pi_j^{BC}\}}\) is used to indicate that 
%it is an interrogated (a measured) variety, and the measurements that we are considering. 
measurements are performed at \(B\), \(C\), and \(BC\).
Also,
the suffixes on the other terms 
%of Eq. (\ref{raat2-to-chhoi}) 
indicate the corresponding measurements that have been performed, 
so that 
\(I(\rho_{A:B|C})_{\{\Pi_i^C,\Pi_j^{BC}\}} = S(\rho_{A|C})_{\{\Pi_i^C\}} - S(\rho_{A|BC})_{\{\Pi_j^{BC}\}}\)
and \(I(\rho_{AB})_{\{\Pi_k^B\}} = S(\rho_A) - S(\rho_{A|B})_{\{\Pi_k^B\}} \equiv S(\rho_A) - \sum_k p_kS(\rho_{A|k})\). 
Optimizing over the measurements, we have the interrogated interaction information for a tripartite state \(\rho_{ABC}\).
%, and is defined as 
%\(I(\rho_{ABC})  = I(\rho_{A:B|C})  - \tilde{I}(\rho_{AB})\). 

 Given a tripartite density operator $\rho_{ABC}$, the interaction 
information \(I(\rho_{ABC})\) is the
 difference between the information shared by the subsystem \(AB\)
when \(C\) is present, and when \(C\) is not present (traced out). 
In some sense, the interaction 
information measures the effect of a bystander \(C\) on the amount of 
correlation shared between \(A\) and \(B\). One can interpret a positive 
interaction information by saying that the presence of \(C\) 
\emph{enhances the correlation} between \(A\) and \(B\). Similarly, negative interaction 
information will mean that the presence of \(C\) somehow \emph{inhibits the correlation}
between \(A\) and \(B\).

Quantum interaction information has the following properties. 
(i) Even though the conditional mutual information and the mutual information 
are both positive, the interaction information can be either positive 
or negative. (ii) It is invariant under local unitaries. (iii) Under unilocal measurements,  
\(I(\rho_{ABC}) \geq \tilde{I}(\rho_{ABC})\). To see (iii), one may check that the identity 
\begin{equation}
 \label{raat-vanish}
I(\rho_{ABC}) - \tilde{I}(\rho_{ABC}) = D(\rho_{AB}) + D(\rho_{BC}) + D(\rho_{CA})
\end{equation}
holds, where in this case, \(I(\rho_{ABC})\) is the optimized version of  \(I(\rho_{ABC})_{\{\Pi_i^A, \Pi_j^B, \Pi_k^C\}} = S(\rho_{A|B})_{\{\Pi_j^B\}} + S(\rho_{B|C})_{\{\Pi_k^C\}} + S(\rho_{C|A})_{\{\Pi_i^A\}} - S(\rho_{ABC})\), 
%(with the optimization being carried 
%out over unilocal measurements)
 and 
\(\tilde{I}(\rho_{ABC}) = \tilde{S}(\rho_{A|B}) + \tilde{S}(\rho_{B|C}) + \tilde{S}(\rho_{C|A}) - S(\rho_{ABC})\), which imply (iii), as 
quantum discord is nonnegative. 
Remarkably, (iii) provides a necessary and sufficient condition for the vanishing 
of quantum discords for the local bipartite states of an arbitrary (pure or mixed) tripartite quantum state. 
As can be seen, discords for the bipartite reduced states vanish if and only if  $\tilde{I}(\rho_{ABC}) = I(\rho_{ABC})$, under unilocal measurements.

%Since the  mutual information and the conditional mutual information 
%involve conditional entropies, in a generalization  to the quantum domain, 
%we need to replace them by specifying the state of one part, given the state 
%of the other. These involve specifying the state of one subsystem, conditional 
%upon measurement on the other subsystem. Accordingly, we will have ``unmeasured 
%interaction information'' and ``interrogated interaction information'' (see below 
%for their definitions). 

\noindent {\bf Theorem I:} For any \(\rho_{ABC}\),
% be a tripartite density operator. 
quantum correlations captured by the bipartite discords will obey monogamy, 
i.e., \(D(\rho_{AB}) + D(\rho_{AC}) \le D(\rho_{A:B C})\) will hold, 
if and only if the interrogated interaction information is less than or equal to
the unmeasured interaction information.

%\begin{proof}
\noindent \emph{Proof.}
It can be checked that if discord respects monogamy then we will have 
\begin{equation}
\label{kal-i-chharbo-ei-paper-ta}
I(\rho_{A:B|C}) - I(\rho_{AB}) \le 
\tilde{I}(\rho_{A:B|C}) - \tilde{I}(\rho_{AB}),
\end{equation}
where the optimizations over measurement bases have already been performed.

Now we prove the converse. 
%We have 
%$I(A:B|\{\Pi_i^C\} ) - I(A:B)_{\{\Pi_i^B\}} \le 
%I(A:B|C) - I(A:B)$. 
%Using 
Assuming (\ref{kal-i-chharbo-ei-paper-ta}), we have
%the relation 
%(\ref{kal-i-chharbo-ei-paper-ta}), 
\(I(\rho_{A:B|C})_{\{\Pi_i^C, \Pi_j^{BC}\}} - I(\rho_{AB})_{\{\Pi_k^B\}} \le 
\tilde{I}(\rho_{A:B|C}) - \tilde{I}(\rho_{AB})\),
and  using the expressions for various mutual informations, we
get
% are led to  
%this leads to 
%\begin{eqnarray}
\(S(\rho_{A|B})_{\{\Pi_k^B\}} + S(\rho_{A|C})_{\{\Pi_i^C\}} - S(\rho_{A|BC})_{\{\Pi_j^{BC}\}}  
% \nonumber\\
\le \tilde{S}(\rho_{A|B}) + \tilde{S}(\rho_{A|C}) - \tilde{S}(\rho_{A|B C})\).
%\end{eqnarray}
%
So we have
%\begin{eqnarray}
\([S(\rho_{A|B})_{\{\Pi_k^B\} } - \tilde{S}(\rho_{A|B}) ] + [S(\rho_{A|C})_{\{\Pi_i^C\}} - \tilde{S}(\rho_{A|C})]
% \nonumber \\
\le  [S(\rho_{A|BC})_{\{\Pi_j^{BC}\}}  - \tilde{S}(\rho_{A|BC})]\).
%.\nonumber \\
%\end{eqnarray}
Consequently, 
%If we minimize this inequality over the measurement bases, 
we 
%will 
have
% the following inequality:
%\begin{eqnarray}
\( [{\min}_{\Pi_i^B} S(\rho_{A|B})_{\{\Pi_i^B \}} - \tilde{S}(\rho_{A|B}) ] 
%\nonumber \\
+ [ {\min}_{\Pi_i^C} S(\rho_{A|C})_{\{\Pi_i^C \}}- \tilde{S}(\rho_{A|C})]  
%\nonumber \\
 \le [ {\min}_{\Pi_i^{BC } } S(\rho_{A|BC})_{\{\Pi_i^{BC}\}}  - \tilde{S}(\rho_{A|BC})]\), leading to (\ref{monogamydiscord}).
%\end{eqnarray}
%Thus, we have 
%\begin{equation}
%D(\rho_{AB}) + D(\rho_{A C}) \le D(\rho_{A:B C}),
%\end{equation}
%when Eq. (\ref{kal-i-chharbo-ei-paper-ta}) is true. 
%Hence, the proof. 
\hfill \Square
%\end{proof}

We note that for any pure tripartite state, the unmeasured interaction information, 
\(\tilde{I}(\rho_{ABC})\),
is zero. As a result, quantum discord is monogamous if and only if 
the interrogated interaction information, \(I(\rho_{ABC})\),  is nonpositive.
As a corollary of Theorem I, we can see that 
%Next, we will show that if we refute condition (2), prove a theorem which can tell us 
%gives us physical insight as to 
quantum discord will be polygamous
%\noindent {\bf Theorem II:} 
if the unmeasured interaction information is negative, i.e., 
\(\tilde{I}(\rho_{ABC}) < (\leq) 0\), and 
the interrogated interaction information is positive, i.e., 
\(I(\rho_{ABC}) \geq (>)  0\).
%, then discord will be 
%polygamous.

%\noindent \emph{Proof.}
% Let \(I(\rho_{ABC}) < 0\). This implies that we have 
%\(\tilde{S}(\rho_{A|B}) + \tilde{S}(\rho_{A|C}) < \tilde{S}(\rho_{A|BC}) + S(\rho_A)\).
%Now let \(I(\rho_{ABC})_{\{\Pi\}} >  0\). This implies that we have 
%\(S(\rho_{A|B})_{\{\Pi_k^B\}} + S(\rho_{A|C})_{\{\Pi_i^C\}} >  S(\rho_{A|BC})_{\{\Pi_j^{BC}\}} +  S(\rho_A)\).
%Upon minimizing on the measurement bases, we have 
%\({\min}_{\Pi_i^B} S(\rho_{A|B})_{\{\Pi_i^B\}} + 
% {\min}_{\Pi_i^C} S(\rho_{A|C})_{\{\Pi_i^C\}} 
%\ge  {\min}_{\Pi_i^{BC } } S(\rho_{A|BC})_{\{\Pi_i^{BC}\}}  +  S(\rho_A)\).
%We therefore have
%\( [{\min}_{\Pi_i^B} S(\rho_{A|B})_{\{\Pi_i^B\}} - \tilde{S}(\rho_{A|B}) ] 
%+ [ {\min}_{\Pi_i^C} S(\rho_{A|C})_{\{\Pi_i^C\}} -\tilde{S}(\rho_{A|C})]  
%> [{\min}_{\Pi_i^{BC } } S(\rho_{A|BC})_{\{\Pi_i^{BC}\}} - \tilde{S}(\rho_{A|B C})]\).
%This leads to a violation of (\ref{monogamydiscord}),
%provided \(I(\rho_{ABC}) < 0\) and 
%\(I(\rho_{ABC})_{\{\Pi\}} >  0\). 
%\hfill \Square

%The above discussion 
This provides a physical insight as to why quantum correlation can 
be polygamous. Suppose that Alice (\(A\)) and Bob (\(B\)) are correlated due to the tripartite 
quantum state shared between \(A\), \(B\), and \(C\). 
Oblivious of the presence of the third observer Charlie (\(C\)), the correlation between Alice and Bob is the quantum mutual information. 
 When there is a third observer (Charlie), the interaction information quantifies the increase in correlations between Alice 
and Bob due to the presence of Charlie. The interaction information can be with or without interrogation on Charlie. 
Our results show that
polygamy is satisfied in the following scenario: (i) When Alice and Bob find out about Charlie's presence, Alice may find that she is more correlated 
with Charlie, before interrogating him, and therefore is in comparison, less correlated with Bob -- mathematically, this is summarized by a negative \(\tilde{I}(\rho_{ABC})\). 
(ii) After interrogating Charlie, Alice finds that she is less correlated with Charlie, but more with Bob  -- this is 
captured by a positive \(I(\rho_{ABC})\).

%interrogation on both Bob (B) and Charlie 
%reduces the correlation between Alice (A) and Bob (interaction information being 
%negative as opposed to unmeasured interaction information being positive) 
%then there the observer Alice does not show much interest in Charlie. She does not try 
%to restrict and wants to get more and more correlated with other observers.
%(Compare this to a girl being interested in more and more boys!).

\emph{Monogamy detects SLOCC-incomparable tripartite classes.--}
We believe that the results obtained on monogamy of quantum discord will be useful in many areas in quantum information. As a first application of 
the results derived, we consider a game where a shopkeeper who sells quantum states promises to provide three-qubit pure states, which are either generalized 
GHZ \cite{GHZ}
%, i.e., of the form  
%\begin{equation}
\(|\psi_{GHZ}\rangle_{ABC} = \cos\Phi |000\rangle + \sin\Phi |111\rangle\),
%\end{equation}
or generalized W states \cite{Wstate, dur-vidal-cirac}
%, i.e., of the form 
%\begin{equation}
\( |\psi_{W}\rangle_{ABC} = \sin\theta\cos\phi |011\rangle + \sin\theta\sin\phi |101\rangle + \cos\theta |110\rangle\).
%\end{equation}
%$\alpha_W=\text{sin}\theta\,\text{cos}\phi,\,\, \beta_W=\text{sin}\theta
%\,\text{sin}\phi$ and $\gamma_W=\text{cos}\theta$.
Here, \(|0\rangle\) and \(|1\rangle\) are orthonormal states of the  corresponding system. 
%While \(\alpha\), \(\beta\), \(\alpha_W\), \(\beta_W\), 
%and \(\gamma_W\) can be considered to be complex numbers, local unitary transformations can make them real, and as such they are 
%henceforth considered to be so. Certainly the normalization conditions 
%\(\alpha^2 + \beta^2 =1\) and \(\alpha_W^2 +\beta_W^2 +\gamma_W^2 =1\) hold.
 The customer is given a large number of copies of the same state. The task is to find out whether the given state is a generalized GHZ or a 
generalized W state.

\begin{figure}[h]%
\includegraphics[width=5cm, height=3.5cm]{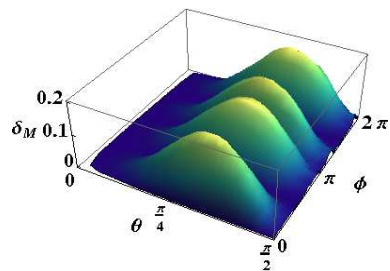}%
\caption{(Color online.) Discord monogamy for generalized W.
 We plot 
\(\delta_M\) (in bits)
%=D(\rho_{AB}) + D(\rho_{AC}) - D(\rho_{A:B C})$ for $|\psi_W\rangle$ on the vertical axis,
 against $(\theta,\, \phi)$ (dimensionless). Here, \(\delta_M (\rho_{ABC}) = D(\rho_{AB}) + D(\rho_{AC}) - D(\rho_{A:BC})\).
%, where $\alpha_W=\text{sin}\theta\,\text{cos}\phi,\,\, \beta_W=\text{sin}\theta
%\,\text{sin}\phi$ and $\gamma_W=\text{cos}\theta$. 
%The base axes are dimensionless while the vertical axis is in bits. 
}%
\label{Fig:wmono}%
\end{figure}

While there does exist other methods to get the answer, we show here that the monogamy inequality in (\ref{monogamydiscord}) can be used 
to give the answer, that additionally provides an interesting perspective of the geometry and structure of tripartite quantum states. 
%
%We now check the status The monogamy ineqaulity,  given in Eq. (\ref{monogamydiscord}) is now checked for two inequivalent class of tripartite states. 
%It was known that the tripartite class of states in which GHZ belongs, known as GHZ class states cannot be obtained from another 
%class of genuinely tripartite entangled state in which W state resides by means of local operations and classical communication (LOCC) 
%with nonzero probability \cite{dur-vidal-cirac}. The generalized GHZ state, given by
%where \(\alpha\) and \(\beta\) are arbitrary complex numbers, satisfying the normalization condition \(|\alpha|^2 + |\beta|^2=1\). 
Among the present cases, the two-party local density matrices of  \(|\psi_{GHZ}\rangle\) are \( \alpha^2 |00\rangle \langle 00| + \beta^2 |11\rangle \langle 11|\), 
whose discord vanishes for all \(\alpha\), so that the monogamy inequality (\ref{monogamydiscord}) 
is always satisfied, as \(D(|\psi_{GHZ})\) is nonnegative for any bipartite partition, and for any \(\alpha\).
%always positive for nonzero \(\alpha\) and \(\beta\). 
%
In sharp contrast, we numerically find that 
the generalized \(W\) states
always violate the monogamy for quantum discord. Numerical simulations were performed for about \(2.5 \times 10^4\) random real sets of \(\{\alpha_W,\beta_W\}\), and 
violation was obtained for all the cases. See Fig. \ref{Fig:wmono}.
% choices of   \(\alpha\), \(\beta\) and \(\gamma\) shows the violation of monogamy relation. 

\begin{figure}[h]%
\includegraphics[width=1.68 in]{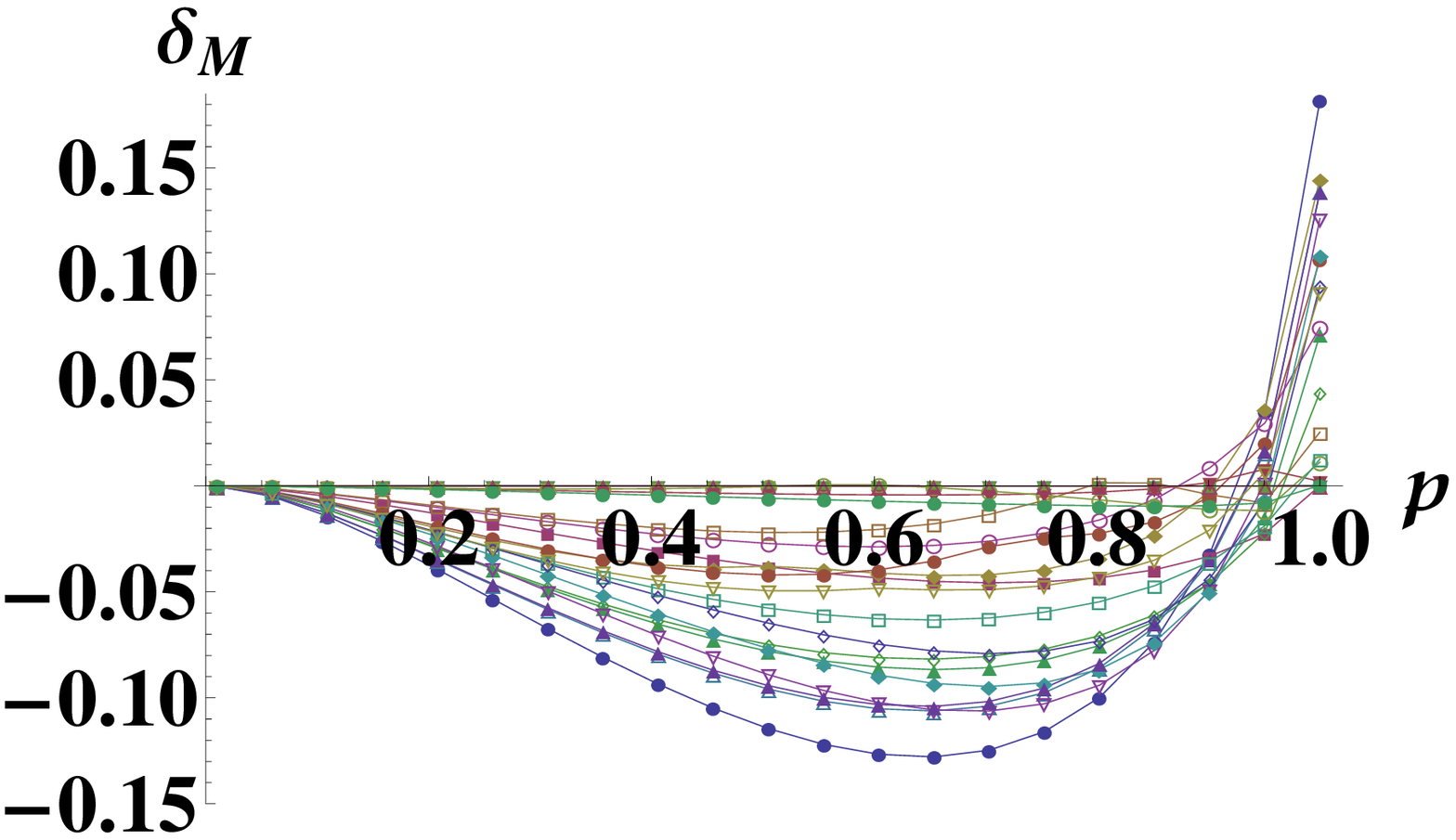}
\includegraphics[width=1.68 in]{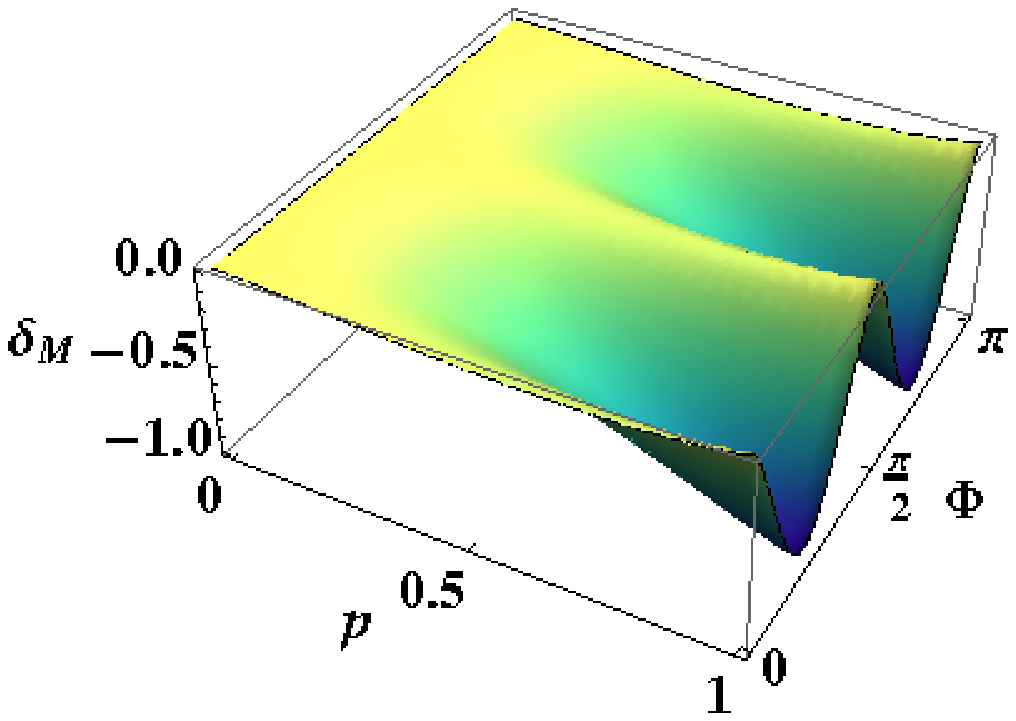}
\caption{(Color online.) Discord monogamy for mixed states. Left: \(\delta_M\) (in bits) against  the mixing parameter \(p\) (dimensionless), 
for different generalized 
\(W\) states. Right: \(\delta_M\) is plotted against \(p\) and \(\alpha\) (dimensionless) for all generalized GHZ states.}
\label{fig:}
\end{figure}

The above result that distinguishes between generalized GHZ and generalized W states is intriguing because of the following fact. Generalized GHZ and 
generalized W states form proper subsets of the GHZ class and W class states. The latter encompasses the whole space of genuine three-party entangled states of 
three-qubit systems.
% \cite{genuine}. 
Moreover, these  two classes form inequivalent classes of three-party states in the sense that they cannot be 
transformed into each other even with stochastic LOCC (SLOCC) maps \cite{dur-vidal-cirac}. 
%
%Therefore, it shows that although quantum discord is a measure of quantum correlations, unlike entanglement, it does not potrait any uniform 
% distribution rule among parties. Apart from that, the above observations of generalized GHZ and \(W\) state leads to a conclusion that monogamy of 
%discord can be used to detect those classes of states. Therefore, it provides another mechanism  to detect these two inequivalent classes by the monogamy relation. 
%
%
%Let us now consider 

An arbitrary (unnormalized) state of the GHZ class is given by 
%\begin{equation}
\(|\psi_G\rangle = \cos \frac{\theta}{2} |000\rangle + |\psi_1\rangle |\psi_2 \rangle |\psi_3 \rangle\),
%\end{equation}
where \(\theta\) is real, and \(|\psi_i\rangle = \alpha_i |0\rangle + \beta_i |1\rangle\), with \(\alpha_i\) and \(\beta_i\) 
being complex numbers satisfying the normalization constraint \(|\alpha_i|^2 + |\beta_i|^2=1\) \((i=1,2,3)\).
% are arbitrary states. 
For 
%these states, 
%we find that satisfying and violation of the monogamy relation are equally likely -- 
\(25000\)
 randomly chosen such states, approximately \(24.49\%\) violate monogamy. 
For an equal number of randomly chosen   W class states, 
violation is obtain for \emph{all} states, indicating that
% from the It is however plausible that the states of the W class would \emph{always} violate monogamy and if so, 
the monogamy relation can 
be used as an indicator for GHZ class states.

Next we ask whether the monogamy (or polygamy) behavior of given pure states is robust against noise admixture. To this end, we 
check the validity of monogamy of quantum discord, for pseudo-pure generalized GHZ and W states, given by 
\(\rho_{GHZ}=(1-p)\mathbb{I}/8 + p|\psi_{GHZ}\rangle \langle \psi_{GHZ}|\) and \(\rho_{W}=(1-p)\mathbb{I}/8 + p|\psi_{W}\rangle \langle \psi_{W}|\), where 
\(0\leq p \leq 1\), and \(\mathbb{I}\) is the identity operator on the three-qubit space.
We perform numerical computations where we need optimizations over arbitrary two-qubit bases, for which we scan over the canonical entangled bases. The 
results are presented in Fig. 2, where we see that the monogamous or polygamous behavior persists for small admixtures of white noise. 
Moreover, admixing white noise to a W class state transforms it from being polygamous to being monogamous. 
This may be seen as due to the monogamous GHZ components present in white noise.
%Admixing white noise to a W class state adds GHZ components to the total state. While W class states are polygamous, the presence of GHZ components may lead to the observed monogamy.

%We also performed numerical computations for GHZ and W class states admixed with white noise. The results 

\emph{Conclusion.--} 
Monogamy is an important aspect of  entanglement, which tells us that this ``costly'' resource is not freely sharable. 
%Here we have raised the question of monogamy for other quantum correlations.
%, such as the quantum discord. 
%Recently, there is an ongoing effort in understanding 
%quantum phenomena by using information-theoretic quantum correlation measures that are independent of the entanglement-separability paradigm. 
We have found that monogamy is not an intrinsic property of other quantum correlation measures.
% like quantum discord.
% may not hold in the general scenario. 
We have delineated necessary and sufficient conditions such that the monogamy can be satisfied. 
Using the notion of interaction information, we have proved that when interrogated interaction information is 
less than or equal to the unmeasured interaction information, then the quantum correlation obeys monogamy. 
Interaction information provides useful insights as to when quantum correlation will be monogamous and polygamous. 
In addition, we find that 
%We then apply 
the monogamy conditions of quantum correlation can be used to distinguish generalized GHZ states from generalized W states.
The results may have applications in quantum information theory and quantum communication tasks. 
%In future, it will be interesting to see if multipartite quantum correlations obey monogamy and if this can 
%detect multipartite inequivalent class of states. 

%\acknowledgments
%\begin{acknowledgments}
We acknowledge computations performed at the cluster computing facility in HRI (http://cluster.hri.res.in/).
%\end{acknowledgments}

\end{document}